\documentclass[12pt]{article}
\usepackage{epsfig,graphicx}
\usepackage{amsmath}
\usepackage{amssymb}
\usepackage{array}
\newcommand{\beq}{\begin{equation}}
\newcommand{\eeq}{\end{equation}}
\newcommand{\bea}{\begin{eqnarray}}
\newcommand{\eea}{\end{eqnarray}}
\newcommand{\no}{\nonumber}

\newcommand{\OMIT}[1]{{}}

\newcommand\spur{\raise.15ex\hbox{/}\kern-.57em }

\newcommand{\CO}{\mathcal{O}}
\newcommand{\cO}{\mathcal{O}}

\newcommand{\CL}{\mathcal{L}}

\newcommand{\vckm}{{V_{\text{CKM}}}}
\newcommand{\upmns}{{\hat U}}
\newcommand{\upmnsdag}{{\hat U^\dagger}}
\newcommand{\upmnstar}{{\hat U^*}}
\newcommand{\XLFV}{\Delta}
\newcommand{\LLN}{{\Lambda_{\rm LN}}}
\newcommand{\LLNsq}{{\Lambda^2_{\rm LN}}}
\newcommand{\LLFV}{{\Lambda_{\rm LFV}}}
\newcommand{\LLFVsq}{{\Lambda^2_{\rm LFV}}}
\newcommand{\hc}{\text{h.c.}}
\newcommand{\lsim}{
\mathrel{\hbox{\rlap{\hbox{\lower4pt\hbox{$\sim$}}}\hbox{$<$}}}}
\newcommand{\gsim}{
\mathrel{\hbox{\rlap{\hbox{\lower4pt\hbox{$\sim$}}}\hbox{$>$}}}}
\def\npb#1#2#3{    {Nucl. Phys.}~B {\bf #1}, #3 (#2)}
\def\plb#1#2#3{    {Phys. Lett.}~B {\bf #1}, #3 (#2)}

\def\prl#1#2#3{    {Phys. Rev. Lett.}~{\bf #1}, #3 (#2)}
\def\prd#1#2#3{    {Phys. Rev.}~D {\bf #1}, #3 (#2)}
\begin{document}

\begin{flushright}
CALT-68-2566\\
UCSD/PTH-05-11\\
June 2005
\end{flushright}
\vspace{1.0 true cm}
\begin{center}
{\Large {\bf  Minimal Flavor Violation in the Lepton Sector} }\\
\vspace{1.0 true cm}
{\large Vincenzo Cirigliano${}^a$, Benjam\'\i{}n Grinstein${}^b$,
Gino Isidori${}^c$, Mark B. Wise${}^a$} \\
\vspace{0.5 true cm}
${}^a$ {\sl California Institute of Technology, Pasadena, CA 91125} \\
\vspace{0.2 true cm}
${}^b$ {\sl Department of Physics, University of California at San Diego, 
La Jolla, CA 92093} \\
\vspace{0.2 true cm}
${}^c$ {\sl INFN, Laboratori Nazionali di Frascati, Via E. Fermi 40, I-00044 Frascati, Italy} \\
\end{center}
\vspace{0.5cm}

\begin{abstract}
We extend the notion of Minimal Flavor Violation  to the lepton sector.
We introduce a symmetry principle
which allows us to express  lepton flavor violation in the
charged lepton sector in terms of neutrino masses and mixing angles. We
explore the dependence of the rates for flavor changing radiative
charged lepton decays ($\ell_i\to\ell_j\gamma$) and $\mu$-to-$e$ conversion in
nuclei on the scales for total lepton number violation, lepton flavor
violation and the neutrino masses and mixing angles. Measurable rates
are obtained when the scale for total lepton number violation  is
much larger than the scale for lepton flavor
violation.

\end{abstract}

\section{Introduction} 
With the discovery of neutrino masses and mixing it has 
been clearly established that lepton flavor is not conserved. 
The smallness of neutrino masses also provides a 
strong indication in favor of the non-conservation 
of total lepton number, although this information 
cannot be directly extracted from data yet.
In analogy to what happens in the quark sector,
the non-conservation of lepton flavor points toward 
the existence of lepton flavor violating (LFV) 
processes with charged leptons, which however have not 
been observed so far. 

Within the Standard Model (SM), the flavor violation in the quark
sector is induced by Yukawa interactions.  These do not break
the baryon number and, to a good approximation, leave the neutral
currents flavor diagonal.  This fact puts very stringent constraints
on the structure of possible new degrees of freedom. New,
flavor-dependent interactions beyond the SM can readily be ruled out
if they contribute significantly to flavor changing neutral currents
(FCNC). One way to evade this constraint is to assume that all the new
degrees of freedom, or at least all the new particles carrying 
flavor quantum numbers, are very heavy. This solution is
theoretically unappealing and largely non-testable.  
A more appealing scenario is to assume that
the new flavor-changing couplings appearing in SM extensions are
suppressed by some symmetry principle.  The most restrictive and
predictive symmetry principle of this type is the so-called Minimal
Flavor Violation (MFV) hypothesis \cite{Georgi,Buras,MFV}: the
assumption that the SM Yukawa couplings are the only sources of
quark-flavor symmetry breaking.  In this case all the cancellations
that render FCNC automatically small in the SM apply just as well to
the new degrees of freedom, allowing for very reasonable mass scales
of the new particles.  Interestingly, the MFV hypothesis can be
formulated in a very general way in terms of an effective field theory
\cite{MFV}, without the need of specifying the nature of the new
degrees of freedom.  In a theory where the new degrees of freedom
also carry  lepton flavor quantum numbers, it is natural to expect that
a similar mechanism occurs also in the lepton sector.

In this paper we extend the notion of MFV to the lepton sector.
In other words, we define and analyze a consistent class of SM 
extensions where the sources of LFV are 
linked in a minimal way to the known structure of the 
neutrino and charged-lepton mass matrices. 
This allows us to address in a general way several interesting questions.
In particular, we shall analyze the general requirements about the 
scale of new physics under which we can expect observable effects 
in low-energy rare LFV processes, such as $\mu \to e\gamma $ 
and $\mu$-to-e conversion in nuclei, without requiring the existence 
of new uncontrollable sources of lepton flavor mixing.  
We shall also  identify some model-independent relations 
among different LFV observables which could allow to falsify 
this general hypothesis about the flavor structure 
of physics beyond the SM.

The large difference between charged lepton and neutrino masses 
is naturally attributed to the breaking of total 
lepton number. This assumption has very 
important consequences to estimate the overall size of 
the LFV terms. As we shall show, only by decoupling the mechanisms
of lepton flavor mixing and lepton number violation can we
generate sizable LFV amplitudes in the charged-lepton sector.
Since lepton flavor and lepton number correspond to two 
independent symmetry groups, this decoupling can  naturally 
be implemented in an effective field theory approach 
with the minimal particle content, namely without introducing 
right-handed neutrino fields. However, in most explicit SM 
extensions this result is achieved by means of the see-saw mechanism 
with heavy right-handed neutrinos. 
For this reason, we shall consider two main possibilities 
in order to define the minimal sources of flavor symmetry breaking 
in the lepton sector: i) a scenario without right-handed neutrinos, 
where the (left-handed) Majorana mass matrix is the only irreducible 
source of  flavor symmetry breaking; ii)  a scenario with right-handed 
neutrinos, where the Yukawa couplings define the irreducible 
sources of flavor symmetry breaking and the (right-handed) Majorana 
mass matrix has a trivial flavor structure.

\section{Minimal breaking of the lepton flavor symmetry}

In the absence of Yukawa couplings,
the flavor symmetry of the quark sector of the SM
would be $SU(3)_Q\times SU(3)_U\times SU(3)_D$ corresponding to individual
rotations of the $Q_{L}^i$, $u_{R}^i$ and $d_{R}^i$ fields 
(the left-handed quark doublet and the two right-handed 
quark singlets) for $i=1,2,3$.  Models with MFV have only two 
independent sources of breaking of this group, namely the two Yukawa couplings 
$\lambda_U$ and  $\lambda_D$. Each of them 
breaks the symmetry in a specific way: in the spurion sense, $\lambda_U$ transforms as
a $(3,\bar 3, 1)$ while $\lambda_D$ as a $(3,1,\bar 3)$. In MFV models any
higher dimension operator that describes long distance remnants of
very short distance physics must be invariant under the full flavor
symmetry group when the couplings $\lambda_U$ and $\lambda_D$ 
are taken to transform as spurions as above \cite{MFV}.

In order to define a similar minimal flavor violating structure 
for the leptons, we first need to specify the field content 
of the theory in the lepton sector. As anticipated, we shall consider two cases:
\begin{enumerate}
\item {\it Minimal field content}: three left-handed 
lepton doublets $L_{L}^i$ and three right-handed charged
lepton singlets $e_{R}^i$ (SM field content). 
In this case the lepton flavor symmetry group is 
\beq
G_{\rm LF} = SU(3)_L\times SU(3)_E~.
\eeq
The lepton sector is also invariant under two $U(1)$ 
symmetries, which can be identified with 
total lepton number, $U(1)_{\rm LN}$, and the weak hypercharge. 
\item {\it Extended field content}: three 
right-handed neutrinos, $\nu^i_R$, in addition to the SM fields.
In this case the field content of the 
lepton sector is very similar to that of the quark sector, 
with a maximal flavor group $G_{\rm LF} \times SU(3)_{\nu_R}$.
\end{enumerate}
In the following we shall define separately the assumptions of 
Minimal Lepton Flavor Violation (MLFV) in these two cases.

\subsection{Minimal Field Content}
In this case the minimal choice for the neutrino mass matrix 
is a left-handed Majorana mass term transforming as $(6,1)$ under $G_{\rm LF}$.
Because of the $SU(2)_L$ gauge symmetry, this mass term cannot
be generated by renormalizable interactions. Moreover, the 
absence of right-handed neutrino fields requires the breaking of
total lepton number. We define the MLFV hypothesis 
in this case as follows:
\begin{enumerate}
\item The breaking of the $U(1)_{\rm LN}$ is independent from the
breaking of the lepton flavor symmetry ($G_{\rm LF}$)
and is associated to a very high scale $\LLN$.
\item There are only two irreducible sources of lepton-flavor
 symmetry breaking, $\lambda_e^{ij}$ and $g_\nu^{ij}$, 
defined by\footnote{~Throughout this paper we use  four-component 
spinor fields, and  $\psi^c=-i\gamma^2\psi^*$ denotes the charge conjugate of the field
 $\psi$. We also use $v = \langle H^0 \rangle \simeq 174$ GeV.}
\begin{align}
\label{eq:minlag}
\CL_{\text{Sym.Br.}} &= - \lambda_e^{ij} \,\bar e^i_R(H^\dagger L^j_L) -\frac1{2 \LLN}\,g_\nu^{ij}(\bar
L^{ci}_L\tau_2 H)(H^T\tau_2L^j_L)+\hc\\
&\to - v \lambda_e^{ij} \,\bar e^i_Re^j_L
- \frac{v^2}{2 \LLN}\,g_\nu^{ij}\,\bar\nu^{ci}_L\nu^j_L+\hc
\end{align}
The smallness of the neutrino mass is attributed to the smallness of
$v/ \LLN$, while  $g_\nu^{ij}$ can have entries of $\CO(1)$ as in the
standard see-saw mechanism. 
\end{enumerate}
The transformation properties of the lepton field under $G_{\rm LF}$
are 
\beq
L_L\to V_L \,L_L~,\qquad\qquad e_R\to V_R \,e_R~.
\eeq
Thus the Lagrangian~(\ref{eq:minlag}) is formally invariant under this
symmetry if the matrices $\lambda_e^{ij}$ and $g_\nu^{ij}$ are taken as
spurions transforming as 
\beq
\label{eq:spurion-rules}
\lambda_e \to V_R^{\phantom{\dagger}} \,\lambda_e V_L^\dagger~, \qquad\qquad
g_\nu \to V_L^{*\phantom{\dagger}} g_\nu V_L^\dagger~.
\eeq

Since we are interested in LFV processes with external charged
leptons, we can use the $G_{\rm LF}$ invariance and rotate the 
fields in the basis where $\lambda_e$ is flavor diagonal. 
In such basis
\bea
\lambda_e &=& \frac{m_\ell }{v}~ = ~\frac{1}{v}\, {\rm diag}(m_e,m_\mu,m_\tau)~, \no \\
g_\nu &=& \frac{\LLN}{v^2} \, \upmnstar  m_\nu  \upmnsdag~ = 
~\frac{\LLN}{v^2} \, \upmnstar {\rm diag}(m_{\nu_1}, m_{\nu_2}, m_{\nu_3} ) \upmnsdag~,
\label{eq:basis}
\eea
where $\upmns$ is the Pontecorvo-Maki-Nakagawa-Sakata (PMNS) mixing matrix.
The latter can be written as $\upmns=U_{e_L}^\dagger U_{\nu_L}$ in terms
of the unitary matrices which connect a generic basis of the lepton 
fields to the mass-eigenstate basis (denoted by a prime):
\beq
\label{eq:mass-rotation}
e_L=U_{e_L}e'_L~,\qquad e_R=U_{e_R}e'_R~,\qquad \nu_L=U_{\nu_L}\nu'_L~.
\eeq
In the basis defined by (\ref{eq:basis}) the simplest 
spurion combination transforming as $(8,1)$ under  $G_{\rm LF}$, 
or the coupling which controls the amount of LFV 
in the charged-lepton sector, is\footnote{~$ g_\nu^\dagger g_\nu$ also contains a
  $(1,1)$ piece under $G_{\rm LF}$. However, it does not contribute to
  lepton flavor violation. Note also that if CP were an exact
  symmetry, $V_L$ in Eq.~(\ref{eq:spurion-rules}) would be required to
  be real, and therefore $ \XLFV_{\rm minimal} =g_\nu$.}
\beq
\label{eq:XLFV}
\left. \XLFV \right|_{\rm minimal} 
 = g_\nu^\dagger g_\nu = \frac{\LLNsq}{v^4}  ~ \upmns  m^2_\nu  \upmnsdag~.
\eeq

\subsection{Extended Field Content}
The second scenario we consider has three 
right-handed neutrinos in addition to the SM fields, 
with a maximal flavor group $G_{\rm LF} \times SU(3)_{\nu_R}$.
There is a large freedom in deciding how to break 
this group  in order to 
generate the observed masses and mixing. In addition to the 
standard Yukawa coupling for the charged leptons, 
in principle we can introduce 
neutrino mass terms transforming as $(6,1,1)$, $(1,1,6)$,
and  $(\bar 3,1,3)$. Since we are interested in 
a minimal scenario, with unambiguous links between 
the irreducible sources of flavor-symmetry breaking and the 
observable couplings in the neutrino mass matrix, we 
must choose only one of these possibilities. 
In order to distinguish this scenario from the previous one, 
and guided by the structure of explicit models with
see-saw mechanism (see e.g.~Ref.~\cite{Borzumati}), 
we make the following assumptions:
\begin{enumerate}
\item 
The right-handed neutrino mass term breaks $SU(3)_{\nu_R}$ to $O(3)_{\nu_R}$, 
namely is proportional to the identity matrix in flavor space:
\beq 
\CL_{\nu_R\text{-mass}}=- \frac{1}{2} M_\nu^{ij}\bar \nu^{ci}_R\nu_R^j+\hc\qquad\text{with}\qquad M_\nu^{ij}=M_\nu
  \delta^{ij}.  
\eeq
\item The right-handed neutrino mass is the only source of $U(1)_{\rm LN}$ 
breaking and its scale is large compared to
 the electroweak symmetry breaking scale: $|M_\nu|\gg v$.
\item The remaining lepton-flavor symmetry is broken only by two
irreducible sources, $\lambda_e^{ij}$ and $\lambda_\nu^{ij}$, defined by
\begin{align}
\label{eq:extlag}
\CL_{\text{Sym.Br.}} 
& = - \lambda_e^{ij} \,\bar e^i_R(H^\dagger L^j_L) +i \lambda_\nu^{ij}\bar\nu_R^i(H^T \tau_2L^j_L)+\hc \no \\
&\to - v\lambda_e^{ij} \,\bar e^i_Re^j_L - v \lambda_\nu^{ij}\,\bar\nu^{i}_R\nu^j_L+\hc
\end{align}
\end{enumerate}
The theory is thus formally invariant under the group 
$G_{\rm LF} \times O(3)_{\nu_R}$, with the following transformation
properties for fields and spurions:
\bea
& L_L\to V_L\, L_L~,\qquad e_R\to V_R\, e_R~, \qquad  \nu_R\to O_\nu\,  \nu_R~,& \no \\
& \lambda_e \to V_R^{\phantom{\dagger}} \,\lambda_e V_L^\dagger~, \qquad
\lambda_\nu \to O_\nu \,\lambda_\nu V_L^\dagger~.&
\label{eq:sp2}
\eea
At low energies the heavy right-handed neutrinos are integrated out, 
generating an effective left-handed Majorana mass matrix,
\beq
m_\nu = \frac{v^2}{M_\nu}\lambda^T_\nu\lambda_\nu~,
\eeq
as in the minimal scenario. Once we identify $M_\nu$ with $\Lambda_{\rm LN}$,
the two scenarios are perfectly equivalent as far as field 
content and operator structure are concerned.  
However, in the extended case the effective left-handed 
neutrino mass matrix  is not an irreducible
source of flavor breaking, since it can be expressed in terms 
of the neutrino Yukawa coupling,~$\lambda_\nu$:
\beq
\frac{v^2 g_\nu}{\Lambda_{\rm LN}} \quad  [{\rm minimal\ case}] \quad 
\leftrightarrow  \quad  
\frac{v^2 }{M_\nu } \lambda_\nu^T\lambda_\nu~
 \quad  [{\rm extended\ case}]
\label{eq:mnu}
\eeq
This has important consequences for LFV processes: the simplest 
spurion combination transforming as $(8,1)$ under $G_{\rm LF}$
is now $\lambda_\nu^\dagger\lambda_\nu$ and is not quadratic
in the neutrino masses as $g_\nu^\dagger g_\nu$ in Eq.~(\ref{eq:XLFV}). 
Since $\lambda^\dagger_\nu\lambda_\nu $ and $\lambda^T_\nu\lambda_\nu $ 
are not necessarily diagonalized by the same orthogonal transformation, 
this also implies that the connection between LFV processes 
and the observables in the neutrino sector is not completely 
unambiguous in this case. We can overcome 
this difficulty by neglecting CP violation in the neutrino mass matrix. 
With this additional assumption,  $\lambda^\dagger_\nu\lambda_\nu$ 
and $\lambda^T_\nu\lambda_\nu $ are the same (real) combination
diagonalized by the (real) PMNS matrix:
\beq
\label{eq:XLFV2}
\left. \XLFV \right|_{\rm extended} 
 = \lambda_\nu^\dagger \lambda_\nu \quad  \stackrel{\rm CP\ limit}{\longrightarrow} \quad
 \frac{M_\nu}{v^2}  ~ \upmns  m_\nu  \upmnsdag~.
\eeq

\section{Operator analysis for LFV processes}
In addition to the SM leptons, we assume that at some scale 
$\LLFV$ above the electroweak scale and well below $\Lambda_{\rm LN}$ (or $M_\nu$)
there are new degrees of freedom carrying lepton flavor
quantum numbers. Integrating them out, their effect will show 
up at low energies as a series of non-renormalizable operators
suppressed by inverse powers of $\LLFV$.
According to the MLFV hypothesis, these operators must be
constructed in terms of SM fields and the spurions $\lambda_e$ and $g_\nu$ (or $\lambda_\nu$),
and must be invariant under $G_{\rm LF}$ when the spurions 
transform as in Eqs.~(\ref{eq:spurion-rules}) or (\ref{eq:sp2}).

We are interested in those operators of dimension five and six that
could lead to LFV process with charged leptons. These operators must 
conserve total lepton number, otherwise they would be suppressed by 
the large $U(1)_{\rm LN}$ breaking scale. As a consequence, no dimension-five 
term turns out to be relevant. For processes involving 
only two lepton fields, such as $\mu \to e\gamma $ 
and $\mu$-to-e conversion, the basic
building blocks are the bilinears $\bar L_L^{i}\Gamma L^j_L$, $\bar e^{i}_R\Gamma L^j_L$
and $\bar e^{i}_R\Gamma e_R^j$. Their indexes must  be contracted with  spurion
combinations transforming under $G_{\rm LF}$ as $(8,1)$, $(\bar 3,3)$  and $(1,8)$, respectively.
Combinations of this type are 
\bea
(8,1) && \qquad  \XLFV,\lambda_e^\dagger\lambda_e,\XLFV^2,\lambda_e^\dagger\lambda_e \XLFV,\ldots  \\
(\bar 3, 3) && \qquad \lambda_e,~\lambda_e \XLFV,~\lambda_e \lambda_e^\dagger \lambda_e~,\ldots \\
(1,8) && \qquad \lambda_e\lambda_e^\dagger,~ \lambda_e \XLFV \lambda_e^\dagger~, \ldots 
\eea 
where $\XLFV$ is defined in Eqs.~(\ref{eq:XLFV}) or (\ref{eq:XLFV2})
for the two scenarios. Given the smallness of $\lambda_e$ (which is
unambiguously fixed by charged lepton masses), we can safely neglect
terms which are of second order in $\lambda_e$.  We shall also assume that
the entries of $\Delta$ are perturbative, retaining only linear terms in
this effective coupling.  In this limit the only relevant LFV
couplings are $\XLFV$ and $\lambda_e \XLFV$. Moreover, we work only to
linear order in the quark Yukawa couplings, $\lambda_U$ and~$\lambda_D$. 

The resulting  dimension-six 
operators bilinear in the lepton fields can be written as 
\begin{equation}
\label{eq:ops-listed}
\begin{aligned} 
O^{(1)}_{LL} &= \bar L_L\gamma^\mu \XLFV L_L\; H^\dagger iD_\mu H\\
O^{(2)}_{LL} &= \bar L_L\gamma^\mu\tau^a \XLFV L_L\; H^\dagger \tau^aiD_\mu H\\
O^{(3)}_{LL} &= \bar L_L\gamma^\mu \XLFV L_L\; \bar Q_L\gamma_\mu Q_L \\
O^{(4d)}_{LL} &= \bar L_L\gamma^\mu \XLFV L_L\; \bar d_R\gamma_\mu d_R \\
O^{(4u)}_{LL} &= \bar L_L\gamma^\mu \XLFV L_L\; \bar u_R\gamma_\mu u_R \\
O^{(5)}_{LL} &= \bar L_L\gamma^\mu\tau^a \XLFV L_L\;\bar Q_L\gamma_\mu\tau^a Q_L 
\end{aligned} 
\qquad
\begin{aligned} 
O^{(1)}_{RL} & = g^\prime 
 H^\dagger\bar e_R\sigma^{\mu\nu} \lambda_e^{\phantom{\dagger}}\XLFV L_L\, B_{\mu\nu}\\
O^{(2)}_{RL} & = g 
 H^\dagger\bar e_R\sigma^{\mu\nu}\tau^a \lambda_e^{\phantom{\dagger}}\XLFV L_L\, W^a_{\mu\nu} \\
O^{(3)}_{RL} & = (D_\mu H)^\dagger \bar e_R \lambda_e^{\phantom{\dagger}}\XLFV D_\mu L_L\\
O^{(4)}_{RL} & =   
 \bar e_R  \lambda_e^{\phantom{\dagger}}\XLFV L_L\, \bar Q_L \lambda_D d_R \\ 
O^{(5)}_{RL} & =  
 \bar e_R \sigma^{\mu\nu} \lambda_e^{\phantom{\dagger}}\XLFV L_L\, 
 \bar Q_L \sigma_{\mu\nu} \lambda_D d_R \\
O^{(6)}_{RL} & =   
 \bar 
e_R  \lambda_e^{\phantom{\dagger}}\XLFV L_L\, \bar u_R  \lambda_U^\dagger  i\tau^2 Q_L \\ 
O^{(7)}_{RL} & =  
 \bar e_R \sigma^{\mu\nu} \lambda_e^{\phantom{\dagger}}\XLFV L_L\, 
 \bar u_R \sigma_{\mu\nu}  \lambda_U^\dagger   i\tau^2 Q_L
\end{aligned} 
\end{equation}  
We have  omitted operators of the type 
$H^\dagger\bar e_R \lambda_e^{\phantom{\dagger}}
 L_L\,H^\dagger H$, which correct the charged lepton mass
matrix but produce no FCNC interactions. 

The operator $O^{(3)}_{RL}$  does not contribute to the
radiative lepton flavor changing decays $\ell_i\to\ell_j\gamma $, and its
contribution to $\mu$-$e$ conversion is suppressed by $m_em_\mu /v^2$. 
The MFV assumption in the quark sector requires the RL operators 
with a quark current to contain
at least one power of the quark Yukawa couplings $\lambda_D$ or $\lambda_U$. Only
the top-quark Yukawa is  non-negligible, and hence, for the low energy
processes we consider $O^{(4)}_{RL}$--$O^{(7)}_{RL} $ can be
neglected. 

Since the top quark Yukawa is order one, in principle, operators
involving higher orders in $\lambda_U$ could be important. They induce
non-negligible FCNC currents in the down-quark sector of the type
$\vckm_{ti}\vckm_{tj}^* \bar d^i_L\gamma^\mu d^j_L$ \cite{MFV}.  For
$\mu$-to-$e$ conversion only the coupling to light quarks is relevant,
and this additional contribution is suppressed by $|\vckm_{td}|^2\ll1$.

\medskip 

In this paper we shall analyze the phenomenological consequences 
of the MLFV hypothesis only in processes involving two lepton 
fields, for which significant prospects of experimental 
improvements are foreseen in the near future \cite{MEG,MECO}.
However, one can in principle apply it also to four-lepton processes,
such as $\mu \to 3 e $.  In this case one needs to extend the operator
basis (\ref{eq:ops-listed}) including the generalization of  $O^{(3-5)}_{LL}$, namely 
\beq
\bar L_L\gamma^\mu \XLFV L_L\; \bar L_L\gamma_\mu L_L~, \qquad 
\bar L_L\gamma^\mu \XLFV L_L\; \bar e_R\gamma_\mu e_R~, \qquad 
\bar L_L\gamma^\mu\tau^a \XLFV L_L\;\bar L_L\gamma_\mu\tau^a L_L~,
\eeq
and also new structures of the type
\beq
\bar L^{c}_L g_\nu L_L  ~  \bar L_L g_\nu^\dagger  L_L^c  
\qquad {\rm or}  \qquad 
\bar L^{c}_L \lambda_\nu^{T} \lambda_\nu L_L  
~  \bar L_L \lambda_\nu^\dagger \lambda_\nu^{*}  L_L^c ~.
\label{eq:opsLc}
\eeq

\subsection{Explicit structure of the LFV couplings}
Given the structure of operators in Eq.~(\ref{eq:ops-listed}),
it is clear that the strength of LFV processes is determined 
by the entries of the matrix $\Delta$ in the mass-eigenstate 
basis of charged leptons. These are listed below for the
two scenarios we are considering, and for the two 
allowed structures (normal and inverted hierarchy)
of the neutrino mass matrix:
\begin{enumerate}
\item {\it Minimal field content}. According to Eq.~(\ref{eq:XLFV}), we have
\beq
\XLFV_{ij} = \frac{\Lambda_{\rm LN}^2}{v^4} \, \left[ 
m_{\nu_1}^{2}  \, \delta_{ij} \, + \, 
\upmns_{i2} \upmns_{j2}^* \, 
\Delta m^2_{\rm sol} 
\pm 
\upmns_{i3} \upmns_{j3}^* \,  \Delta m^2_{\rm atm} \right]~, 
\eeq
where $\Delta m^2_{\rm atm}$ and $\Delta m^2_{\rm sol}$ denote the  
squared mass differences deduced from atmospheric and solar neutrino data, 
respectively. 
The plus sign corresponds to normal hierarchy ($m_{\nu_1} < m_{\nu_2} \ll m_{\nu_3}$), 
while the minus one to the inverted case 
($m_{\nu_3} \ll  m_{\nu_1} < m_{\nu_2}$). 
Explicitly, using the PDG notation of the PMNS matrix  
(we adopt the convention that $s_{13}\geq0$ and $0\leq\delta<2\pi $)
 \cite{PDG},
we find 
\bea
\XLFV_{\mu e} & = & 
\frac{\Lambda_{\rm LN}^2}{v^4} \, \frac{1}{\sqrt{2}} \, \left(
s \, c  \, \Delta m^2_{\rm sol} 
\pm s_{13} \, e^{i \delta} \, \Delta m^2_{\rm atm} 
\right) 
\equiv \frac{\Lambda_{\rm LN}^2}{v^2} \, a_{\mu e}~, 
\nonumber \\
\XLFV_{\tau e} & = & 
\frac{\Lambda_{\rm LN}^2}{v^4} \, \frac{1}{\sqrt{2}} \, \left(
- s \, c  \, \Delta m^2_{\rm sol} 
\pm s_{13} \, e^{i \delta} \, \Delta m^2_{\rm atm} 
\right)
\equiv \frac{\Lambda_{\rm LN}^2}{v^2} \, a_{\tau e}~, 
\nonumber \\
\XLFV_{\tau \mu} & = & 
\frac{\Lambda_{\rm LN}^2}{v^4} \, \frac{1}{2} \, \left(
-  c^2  \, \Delta m^2_{\rm sol} 
\pm  \Delta m^2_{\rm atm} 
\right)
\equiv \frac{\Lambda_{\rm LN}^2}{v^2} \, a_{\tau \mu}~,  
\label{eq:aij}
\eea
where we have assumed maximal mixing for the atmospheric 
case and $s$ and $c$ denote sine and cosine of the solar mixing angle. 
In a given scenario for the spectrum (normal or inverted), 
the dimensionless couplings $a_{i j}$ are completely fixed by  
oscillation experiments modulo the dependence on the 
combination $s_{13} e^{i \delta}$. 

\item{} {\em Extended field content}. According to
  Eq.~(\ref{eq:XLFV2}), assuming CP conservation in the lepton sector we have
\beq
\XLFV_{ij} =
\frac{M_\nu}{v^2} \, \left[ 
m_{\nu_1} \, \delta_{ij} \, + \, 
\upmns_{i2} \upmns_{j2} \, 
(m_{\nu_2} - m_{\nu_1})  
+  
\upmns_{i3} \upmns_{j3} \,  
(m_{\nu_3} - m_{\nu_1}) \right]~.
\eeq
Note that the assumption of 
CP conservation forces us to choose the PMNS phase 
$\delta = 0$ or $\pi$. 
Hence,
\bea
\XLFV_{\mu e} & = & 
\frac{M_\nu}{v^2} 
\, \frac{1}{\sqrt{2}} \, \left[
s \, c  \, 
(m_{\nu_2} - m_{\nu_1})  
\pm s_{13} \, %%e^{i \delta} \, 
(m_{\nu_3} - m_{\nu_1})   
\right]
\equiv \frac{M_\nu}{v}  \, b_{\mu e}~,  
\nonumber \\
\XLFV_{\tau e} & = & 
\frac{M_\nu}{v^2} \, \frac{1}{\sqrt{2}} \, \left[
- s \, c  \, 
(m_{\nu_2} - m_{\nu_1})  
\pm s_{13} \, %%e^{i \delta} \, 
(m_{\nu_3} - m_{\nu_1})   
\right]
\equiv \frac{M_\nu}{v}  \, b_{\tau e}~,  
\nonumber \\
\XLFV_{\tau \mu} & = & 
\frac{M_\nu}{v^2} \, \frac{1}{2} \, \left[
-  c^2  \, 
(m_{\nu_2} - m_{\nu_1})  
+  
(m_{\nu_3} - m_{\nu_1})   \right]
\equiv \frac{M_\nu}{v}  \, b_{\tau \mu}~,
\label{eq:bij}
\eea
where the $+$ and $-$ signs correspond to $\delta=0$ and $\pi$,
respectively. 
In the  normal hierarchy case ($\nu_1$ is the lightest neutrino), one has:
\beq
m_{\nu_2} - m_{\nu_1}  ~ \stackrel{m_{\nu_1} \to 0}{\longrightarrow} ~
\sqrt{\Delta m^{2}_{\rm sol}}~,  \qquad\quad  
m_{\nu_3} - m_{\nu_1}  ~ \stackrel{m_{\nu_1} \to 0}{\longrightarrow} ~
\sqrt{\Delta m^{2}_{\rm atm}}~,
\eeq
while in the inverted hierarchy case ($\nu_3$ is the lightest neutrino)
\beq
m_{\nu_2} - m_{\nu_1} ~ \stackrel{m_{\nu_3} \to 0}{\longrightarrow} ~
\frac{\Delta m^2_{\rm sol}}{2 \sqrt{\Delta m^{2}_{\rm atm}}}~,   \qquad\quad  
m_{\nu_3} - m_{\nu_1} ~  \ \stackrel{m_{\nu_3} \to 0}{\longrightarrow} ~
 -  \sqrt{\Delta m^{2}_{\rm atm}}~.
\eeq
\end{enumerate} 
After using input from oscillation experiments, 
the couplings $b_{i j}$ still depend on the spectrum ordering,
the lightest neutrino mass, and the value of $s_{13}$
(the dependence from $\delta$ has disappeared because of the 
assumption of CP conservation).

\section{Phenomenology}
We are now ready to analyze the phenomenological implications 
of the new LFV operators. In particular, we are interested in answering 
the following questions: (i) under which conditions on the new physics scales
$\Lambda_{\rm LN}$ (or $M_\nu$) and $\Lambda_{\rm LFV}$ can we expect observable
effects in low energy reactions and therefore positive signals in
forthcoming experiments?  (ii) is there a specific pattern in the
decay rates predicted by MLFV?  Can we use it to falsify the
assumption of minimal flavor violation?

In order to address these issues, we will study the rates 
for $\mu \to e$ conversion in nuclei $\Gamma^{A}_{\rm conv} \equiv \Gamma (\mu^- +
A(N,Z) \to e^- + A(N,Z))$, experimentally normalized to the capture
rate $ \Gamma^{A}_{\rm capt} \equiv \Gamma (\mu^- +A(Z,N)\to \nu_\mu+ A(Z-1,
N+1))$,  and the radiative decays $\mu \to e \gamma$, $\tau \to \mu \gamma$, 
$\tau \to e \gamma$. Throughout, we  will use normalized 
branching fractions defined as:
\beq
B_{\mu\to e}^{A}\equiv \frac{\Gamma^{A}_{\rm conv}}{ \Gamma^{A}_{\rm capt} }~,
\qquad\qquad
B_{\ell_i \to \ell_j \gamma} \equiv 
\frac{ \Gamma (\ell_i \to \ell_j \gamma) }{
\Gamma (\ell_{i} \to \ell_{j} \nu_{i} \bar{\nu}_{j} )} \ . 
\label{eq:lagrLFV}
\eeq
The starting point of our analysis is the effective Lagrangian
generated at a scale $\LLFV$ 
\beq
\label{eq:eff-lag}
\CL=\frac{1}{\LLFVsq}
\sum_{i=1}^5 c_{LL}^{(i)} 
O_{LL}^{(i)} +   \frac{1}{\LLFVsq} \left(
\sum_{j=1}^2 c_{RL}^{(j)} O_{RL}^{(j)}  +\hc  \right)
\eeq
In principle one should evolve this Lagrangian 
down to the mass of the decaying particles.
However, for the purpose of the present work we 
shall neglect the effect of electroweak corrections
and treat the $c^{(i)}$ as effective 
Wilson coefficients renormalized at the low scale.  
In terms of these we find 
(in the limit $m_{\ell_j} \ll  m_{\ell_i}$) 
\beq
B_{\ell_i \to \ell_j \gamma} = 
384 \pi^2 e^2   \, \frac{v^4}{\Lambda_{\rm LFV}^4} \, 
\left| \Delta_{ij} \right|^2   
\,  \left| c_{RL}^{(2)} -  c_{RL}^{(1)} \right|^2  
\label{eq:Br1}
\eeq
and 
\begin{eqnarray}
B_{\mu \to e}^{A} &=& 
\frac{32 \, G_F^2 \, m_\mu^5}{\Gamma_{\rm capt}^A} \, 
\frac{v^4}{\Lambda_{\rm LFV}^4} \, 
\left| \Delta_{\mu e} \right|^2 \  
\Bigg|
 \left( \Big( \frac14-s_w^2 \Big) V^{(p)}-\frac14V^{(n)} \right) 
\, (c_{LL}^{(1)}+c_{LL}^{(2)}) 
\nonumber \\
&+& 
\frac{3}{2} (V^{(p)}+V^{(n)}) \, c_{LL}^{(3)} +
(V^{(p)}+\frac{1}{2}V^{(n)}) \, c_{LL}^{(4u)} +
(\frac{1}{2}V^{(p)}+V^{(n)}) \, c_{LL}^{(4d)} 
\nonumber \\ 
&+& \frac{1}{2}(-V^{(p)}+V^{(n)}) \, c_{LL}^{(5)}  
- \frac{e D}{4} ( c_{RL}^{(2)} -  c_{RL}^{(1)} )^{*}  
\Bigg|^2 \ , 
\label{eq:Br2}
\end{eqnarray}
where we use the notation of Ref.~\cite{Kitano:2002mt}
for the dimensionless  
nucleus-dependent overlap integrals $V^{(n)}$, $V^{(p)}$, $D$ 
and we denote by  $s_w = \sin \theta_w = 0.23$ the weak mixing angle.

\subsection{Minimal field content}
Let us now consider the scenario with minimal field content. 
By making in Eqs.~(\ref{eq:Br1}) and ~(\ref{eq:Br2}) the replacement
\beq 
\frac{v^4}{\Lambda_{\rm LFV}^4} \,  \left|\Delta_{ij}\right|^2  \longrightarrow 
\frac{\Lambda_{\rm LN}^4}{\Lambda_{\rm LFV}^4} \, \left|a_{ij}\right|^2  \ , 
\eeq
one can see that all LFV rates have the following structure
\beq
B_{\ell_i \to \ell_j (\gamma)} = 10^{-50}
\left(\frac{\Lambda_{\rm LN}}{\Lambda_{\rm LFV}}\right)^4
R_{\ell_i \to \ell_j (\gamma)} (s_{13}, \delta; c^{(i)})~.
\label{eq:Ri}
\eeq
The overall numerical factor $10^{-50}$ is chosen such 
that the $R_{\ell_i \to \ell_j (\gamma)}$ have a natural 
size of $\cO(1)$. Its value can easily be understood by 
noting that 
$|a_{ij}|^2 \lsim (\Delta m^2_{\rm atm}/v^2)^2 \approx 10^{-52}$.

\begin{table}[t]
\begin{center}
\begin{tabular}{|c|c|c|c|} \hline 
\raisebox{3pt}[15pt][5pt]{$\Delta m^2_{\rm sol}$} & 
\raisebox{3pt}[15pt][5pt]{$\Delta m^2_{\rm atm}$} & 
\raisebox{3pt}[15pt][5pt]{$\theta_{\rm sol}$} &  
\raisebox{3pt}[15pt][5pt]{$s_{13}^{\rm max}$}   \\ \hline 
\raisebox{0pt}[15pt][5pt]{$8.0\times 10^{-5}~{\rm eV}^2$} &
\raisebox{0pt}[15pt][5pt]{$2.5\times 10^{-3}~{\rm eV}^2$} & 
\raisebox{0pt}[15pt][5pt]{$33^\circ $} & 
\raisebox{0pt}[15pt][5pt]{$0.25$} \\ \hline
\end{tabular}
\end{center}
\caption{\label{tab:inputs}
Reference values of neutrino mixing parameters 
used in the phenomenological analysis 
(for a detailed discussion see e.g.~\cite{strumia}).}
\end{table}

\begin{figure}[!t]
\centering
\epsfxsize=8cm
\epsffile{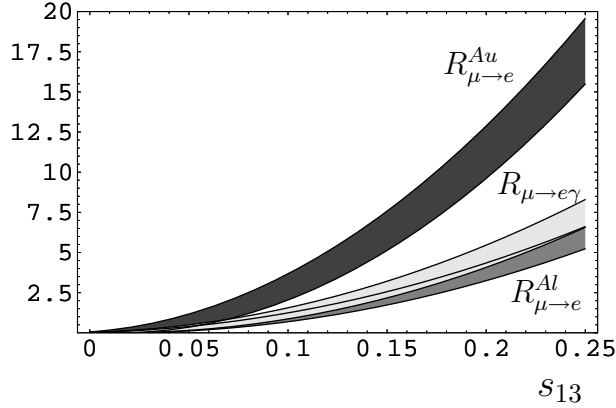}
\caption{\label{fig:Ri} 
Ratios $R_{i}$ defined in Eq.~(\ref{eq:Ri}) 
as a function of $s_{13}$  for $c^{(3)}_{LL} = c^{(2)}_{RL} = 1$ and 
all other $c^{(i)} = 0$. 
The shaded bands correspond to variation of the phase $\delta$ between 
$0$~and~$\pi$. }
\end{figure}

\begin{figure}[t]
\centering
\epsfxsize=15cm
\epsffile{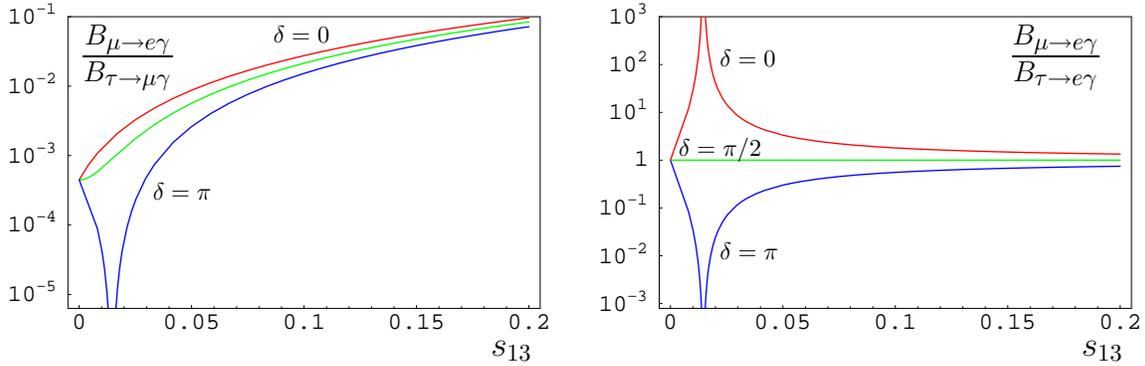}
\caption{\label{fig:ratios1} 
Ratios $B_{\mu \to e \gamma} /B_{\tau \to \mu \gamma}$ (left) and
$B_{\mu \to e \gamma} /B_{\tau \to e \gamma}$ (right) as a function of
$s_{13}$ for different values of the CP violating phase $\delta$  
in the normal hierarchy case. The uncertainty due to the first 3 
entries in table~\ref{tab:inputs} is not shown.}

\end{figure}

\begin{figure}[t]
\centering
\vskip 0.5 cm 
\epsfxsize=8cm
\epsffile{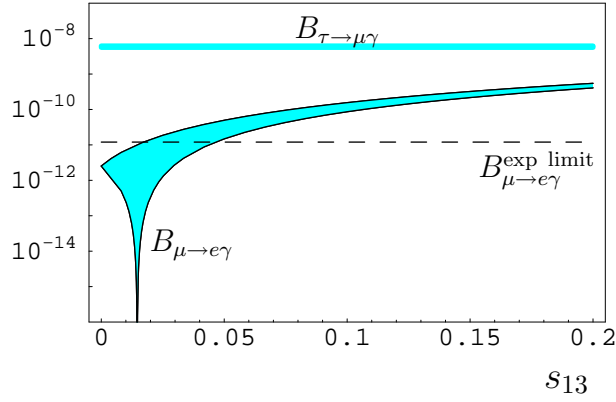}
\caption{\label{fig:tau} 
$B_{\tau \to  \mu \gamma}$ and 
$B_{\mu  \to  e \gamma}$ as a function of 
$s_{13}$, for $\LLN/\LLFV = 10^{10}$ and 
$c^{(2)}_{RL} - c^{(1)}_{RL}= 1$. 
The shading corresponds to different values of the phase $\delta$ 
and the normal/inverted spectrum.
The uncertainty due to the first 3 
entries in table~\ref{tab:inputs} is not shown. }
\end{figure}

A glance at the explicit
structure of the $a_{ij}$ in Eq.~(\ref{eq:aij}) shows that their size
is maximized for $s_{13} = s_{13}^{\rm max}$ (in both normal and
inverted hierarchy), due to $\Delta m^2_{\rm atm} \gg \Delta m^2_{\rm
sol}$.  In order to derive order-of-magnitude conditions on the ratio
$\Lambda_{\rm LN}/\Lambda_{\rm LFV}$, we consider the reference case 
defined by $s_{13} = s_{13}^{\rm max}$, $\delta = 0$, and the reference 
values quoted in Table~\ref{tab:inputs}. Then setting all the 
Wilson coefficients to zero but for $c_{RL}^{(2)}= c_{LL}^{(3)}=1$,
and using the overlap integrals and capture rates reported in 
Ref.~\cite{Kitano:2002mt} (table I of \cite{Kitano:2002mt}), we find
\beq
B_{\mu \to e} ~=~ \left(\frac{\LLN}{\LLFV}\right)^4 \begin{cases}
6.6 \times10^{-50}& \text{for Al} \\
19.6 \times10^{-50}&\text{for Au}
\end{cases}
\qquad \qquad 
B_{\mu \to e \gamma} ~=~  8.3 
\times 10^{-50}~\left(\frac{\LLN}{\LLFV}\right)^4~.
\label{eq:muec1}
\eeq 
Despite the strong dependence of the numerical coefficients in
Eq.~(\ref{eq:muec1}) on $s_{13}$, 
illustrated in Figure~\ref{fig:Ri}, 
these results allow us to draw several  interesting conclusions.
\begin{itemize}
\item
If there is no large hierarchy 
between the scales of lepton-number and lepton-flavor violation, 
there is no hope to observe LFV signals in charged-lepton 
processes. On the other hand, if $\Lambda_{\rm LFV}$ is not far 
from the TeV scale (as expected in many realistic scenarios),
it is natural to expect visible LFV processes 
for a wide range of $\Lambda_{\rm LN}$: from $10^{13}$~GeV
up to the GUT scale. For instance a $B_{\mu \to e} = \cO(10^{-13})$, 
within reach of the MECO experiment, is naturally 
obtained for $\LLN \sim 10^9 \LLFV$, which for $\LLFV \sim  10~\text{TeV}$ 
implies $\LLN \sim 10^{13}~\text{GeV}$.
Such a ratio of scales would also imply $B_{\mu \to e \gamma}=\cO(10^{-13})$, 
within the reach of the MEG experiment.
\par
Note that  the requirement of ``perturbative'' 
treatment of the couplings $g_{\nu}$, together with upper 
limits on the light neutrino masses, implies upper limits on 
the scale $\LLN \simeq v^2 g_\nu/m_\nu$. 
By loosely requiring $|g_{\nu}|<1$ one obtains 
$\LLN \lsim  3\times 10^{13}\, (1 \, {\rm eV}/m_{\nu}) \,  {\rm GeV}$. 
This means that we cannot make the ratio $\LLN/\LLFV$ arbitrarily 
large. 
\item
Interestingly,  
$\mu \to e$ conversion and $\mu \to e \gamma$
have a quite different sensitivity on the type 
of operators involved. In particular, while 
$\mu \to e \gamma$ is sensitive only to the LR operators,
the $\mu \to e$ conversion is more sensitive 
to the LL terms:
\bea
\frac{ B_{\mu \to e}(c_{RL}^{(2)}=1,~{\rm other}~c^{(i)}=0) }
{ B_{\mu \to e}(c_{LL}^{(3)}=1,~{\rm other}~c^{(i)}=0) } &=&
\begin{cases}
3 \times 10^{-3} & \text{for Al}~, \\
1.5 \times10^{-3} & \text{for Au}~,
\end{cases} \\
\frac{ B_{\mu \to e}(c_{RL}^{(2)}=1,~{\rm other}~c^{(i)}=0) }
{ B_{\mu \to e}(c_{LL}^{(1)}=1,~{\rm other}~c^{(i)}=0) } &=&
\begin{cases}
0.47   & \text{for Al}~,\\
0.17   & \text{for Au}~.
\end{cases}
\eea

\item
The comparison of the various $B_{l_i \to l_j \gamma}$ rates is a
useful tool to illustrate the predictive power of the MLFV (and
eventually to rule it out from data). In Figure~\ref{fig:ratios1} we
report the ratios $B_{\mu \to e \gamma} /B_{\tau \to \mu \gamma}$ and
$B_{\mu \to e \gamma} /B_{\tau \to e \gamma}$ as a function of
$s_{13}$ for three different values of the CP violating phase
$\delta$ in the normal hierarchy case (the inverted case is obtained
by replacing $\delta$ with $\pi - \delta$).  One observes the clear
pattern $B_{\tau \to \mu \gamma} \gg B_{\tau \to e \gamma} \sim B_{\mu
\to e \gamma}$, with hierarchy increasing as $s_{13} \to 0$.
Observation of deviations from this pattern could in the future
falsify the hypothesis of minimal flavor violation in the lepton
sector. 
\item
There is a window in parameter space where we can expect observable
effects in $\tau$ decays. As illustrated in Figure~\ref{fig:tau},
$B_{\tau \to \mu \gamma}$ does
not depend on $s_{13}$, while $B_{\mu e}$ does.  In the normal
hierarchy case, for $\delta =\pi$ and $s_{13} \to s c \Delta m^2_{\rm
sol}/\Delta m^2_{\rm atm}$, one has $B_{\mu \to e \gamma} \to 0$.
Therefore, close to this region of parameter space one can have 
a sizable $B_{\tau \to \mu \gamma}$ while $B_{\mu \to e\gamma}$ 
can be kept below the present experimental limits. 
In particular, for $\LLN \sim 10^{10} \, \LLFV$ 
we find $B_{\tau \to \mu \gamma} \sim 10^{-8}$,
which implies a branching ratio for $\tau \to \mu \gamma$
above $10^{-9}$ possibly observable at (super) $B$ factories.
Note that a change in the ratio 
$\LLN/\LLFV$ would only result into a shift 
of the vertical scale in Figure~\ref{fig:tau},
without affecting the relative distance between the 
$B_{\mu \to e \gamma}$ and $B_{\tau \to \mu \gamma}$
bands.
\end{itemize}


\subsection{Extended field content}
\begin{figure}[t]
\centering
\epsfxsize=15cm
\epsffile{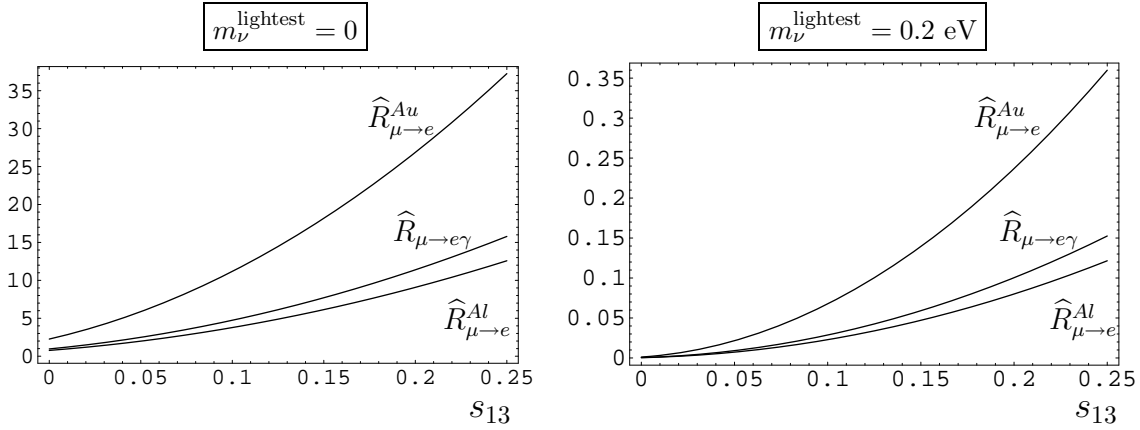}
\caption{\label{fig:n1} 
Ratios $\widehat R_{i}$ defined in Eq.~(\ref{eq:Ri2}) 
as a function of $s_{13}$  for $c^{(3)}_{LL} = c^{(2)}_{RL} = 1$,  
all other $c^{(i)} = 0$, normal spectrum and $\delta=0$.}
\end{figure}

\begin{figure}[t]
\centering
\epsfxsize=15cm
\epsffile{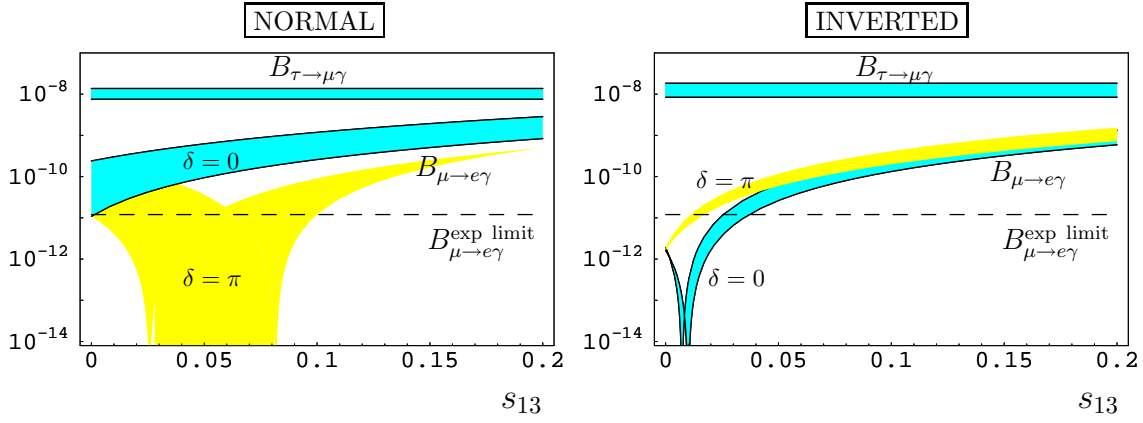}
\caption{\label{fig:last} 
$B_{\tau \to  \mu \gamma}$ and $B_{\mu  \to  e \gamma}$ as a function of 
$s_{13}$, for $(v M_\nu) /\LLFVsq = 5 \times 10^{7}$ and 
$c^{(2)}_{RL}-c^{(1)}_{RL}=1$.   
The shading corresponds to different values of the lightest neutrino 
mass, ranging between 0 and 0.02~eV. The two choices of $\delta$
correspond to the $\pm$ sign in Eq.~(\ref{eq:bij}). }
\end{figure}

The discussion of the extended model proceeds in a very similar way,
by replacing the dimensionless couplings $a_{ij}$ with the $b_{ij}$,
and $(\LLN/\LLFV)^4$ with $(vM_\nu/ \LLFVsq)^2$. The analog of 
Eq.~(\ref{eq:Ri}) reads
\beq
B_{\ell_i \to \ell_j (\gamma)} = 10^{-25}
\left(\frac{v M_\nu }{\Lambda^2_{\rm LFV}}\right)^2
\widehat R_{\ell_i \to \ell_j (\gamma)} (s_{13}, m_{\nu}^{\rm lightest}; c^{(i)})~.
\label{eq:Ri2}
\eeq
As illustrated in Figure~\ref{fig:n1},
the dimensionless functions $\widehat R_{\ell_i \to \ell_j (\gamma)}$
depend on both $s_{13}$ and $m_{\nu_{\rm lightest}}$, 
and the maximal values are obtained 
for $s_{13} = s_{13}^{\rm max}$ and $m_{\nu_{\rm lightest}} \to 0$.
In order to explore the sensitivity to the scale ratio 
$v M_\nu/\Lambda^2_{\rm LFV}$, we again pick a favorable reference point ($s_{13} =
s_{13}^{\rm max}$ and $m_{\nu_{\rm lightest}} = 0$) and set all the
Wilson coefficients to zero except for $c_{RL}^{(2)}= c_{LL}^{(3)}=1$.
In the normal hierarchy case we then find
\beq
B_{\mu \to e} ~=~ \
\left(\frac{vM_\nu}{\LLFVsq}\right)^2  \begin{cases}
1.3 \times10^{-24}& \text{for Al} \\
3.7 \times10^{-24}&\text{for Au}
\end{cases}
\qquad  
B_{\mu \to e \gamma} ~=~  1.6 
\times 10^{-24}~\left(\frac{vM_\nu}{\LLFVsq}\right)^2 
\label{eq:muec2}
\eeq 
while for the inverted case:
\beq
B_{\mu \to e} ~=~ \
\left(\frac{vM_\nu}{\LLFVsq}\right)^2  \begin{cases}
6.7 \times10^{-25}& \text{for Al} \\
2.0 \times10^{-24}&\text{for Au}
\end{cases}
\qquad  
B_{\mu \to e \gamma} ~=~  8 
\times 10^{-25}~\left(\frac{vM_\nu}{\LLFVsq}\right)^2 
\label{eq:muec3}
\eeq 
The general conclusions we can infer from this scenario are the 
following:
\begin{itemize}
\item
As was the case with minimal field content, a large hierarchy between
$M_\nu$ and $\LLFV$ is required to obtain observable effects.  For
example, fractions $B_{\mu\to e(\gamma)}=\cO(10^{-13})$ are obtained for $M_\nu \sim
3 \times 10^5 \LLFVsq/v$, which for $\LLFV \sim 10~\text{TeV}$ gives $M_\nu \sim 2\times
10^{11}~\text{GeV}$.
\item
Comparing the values of $M_\nu$ in this scenario 
versus the corresponding values of $\LLN$ 
in the minimal case, we find that the same effect 
in a given LFV process is typically obtained for 
$\LLN > M_\nu$. This can be understood by 
noting that in the minimal case the LFV amplitudes 
are proportional to combinations 
of the type $\Delta m^2_\nu \LLN^2/v^4$, 
while in the extended case this factor is
replaced by $\Delta m_\nu M_\nu /v^2$. 
The two scales of lepton number violation 
are indeed equivalent when this factor 
is of $\cO(1)$, namely for the maximal 
value allowed by the ``perturbative condition'' 
on the Yukawa couplings: 
$M_\nu \sim \LLN \sim v^2/\Delta m_\nu \sim 10^{15}$~GeV.
\item
Much like in the previous scenario, in this case the ratios of the
various LFV rates are unambiguously determined in terms of neutrino
masses and mixing angles; however, the results are potentially
different than in the minimal case because of the different relation
between LFV parameters and neutrino mass matrix. As illustrated in
Fig.~\ref{fig:last}, one still observes the pattern $B_{\tau \to \mu \gamma} \gg
B_{\mu\to e \gamma}$ ($\sim B_{\tau \to e \gamma} $).  For a given choice of $\delta=0$ or $\pi$,
the strength of the $\mu\to e$ suppression is very sensitive to whether
the hierarchy is normal or inverted.  For $\delta = 0$ the present
experimental limit on $B_{\mu \to e \gamma}$ allows large values of $B_{\tau \to \mu
\gamma}$ only for the inverted hierarchy, whereas for $\delta = \pi$, a large
region with a sizable $B_{\tau \to \mu \gamma}$ is allowed only for the normal
hierarchy. Note that the overall vertical scale of Fig.~\ref{fig:last}
depends on both the ratio  $(v M_\nu) /\LLFVsq$ -- as implied by Eq.~\ref{eq:Ri2} --
and the value of the lightest neutrino mass (as illustrated by Fig.~\ref{fig:n1}). 
\end{itemize}

\section{Conclusions}
In this paper we extend the notion of MFV to the lepton sector.  
We define a symmetry principle and an effective field theory 
which allows us to relate, in a general way, 
lepton-flavor mixing in the neutrino sector 
to lepton-flavor violation in the charged lepton sector.
The construction of such effective theory allow us to address 
fundamental questions about the flavor structure of the lepton 
sector in a very general way, insensitive to many 
details about the physics beyond the SM.
 
We find two  ways in which we can define the sources of flavor symmetry breaking 
in the lepton sector in a minimal and thus very predictive way:
i) a scenario where the left-handed Majorana mass matrix is the 
only irreducible source of flavor symmetry breaking; 
and ii) a scenario with heavy right-handed neutrinos, 
where the Yukawa couplings define the irreducible 
sources of flavor symmetry breaking and the right-handed Majorana 
mass matrix has a trivial flavor structure.
 
We find that visible LFV effects in the charged lepton sector
are generated when there is a large hierarchy between the scales 
of lepton flavor mixing ($\LLFV$) and total lepton number violation ($\LLN$).
This condition is indeed realized within the explicit 
extensions of the SM widely discussed in the literature 
which predict sizable LFV effects in charged leptons
\cite{Borzumati,Barbieri} (for an updated discussion 
see Ref.~\cite{review} and references therein).
Within our general framework, we find that the 
new generation of experiments on LFV would naturally probe 
the existence of new degrees of freedom carrying 
lepton flavor numbers up to 
energy scales of the order of $10^3$~TeV, if the scale 
$\LLN$ is close to the GUT scale.
 
The two scenarios we have considered are highly predictive and
possibly testable by future experiments. While the rates for LFV
effects strongly depend on the (unknown) scales $\LLN$ and $\LLFV$,
the ratio of different LFV rates are unambiguously predicted in terms
of neutrino masses and mixing angles.  At present, the uncertainty in
our predictions for such ratios arises mainly from the poorly
constrained value of $s_{13}$ and, to a lesser extent, from the neutrino
spectrum ordering and the CP violating phase $\delta$. 

One of the  clearest consequences from the phenomenological
point of view is that if $s_{13} \gsim 0.1$ there is no 
hope to observe $\tau \to \mu \gamma$ at future accelerators.
On the other hand, $\mu \to e \gamma$ and  $\mu$-to-$e$ conversion
are within reach of future experiments for reasonable
values of the symmetry breaking scales.

\bigskip

\paragraph{Acknowledgments}
We thank Curtis Callan for valuable discussions. 
V.C.~was supported by Caltech through the Sherman Fairchild fund, 
and acknowledges the hospitality of the LNF Spring Institute 2005. 
This work was supported by the U.S.~Deartment of Energy  
under grants DE-FG03-97ER40546 and DE-FG03-92ER40701.

\newpage

\section*{Appendix}

\setlength{\extrarowheight}{2pt}
\begin{table}[ht]
\begin{center}
\begin{tabular}{|c|c|c|c|c|} \hline 
& $V^{(p)}$ & $V^{(n)}$ & $D$ & $\Gamma_{\rm capt}$ ($10^6\text{s}^{-1}$)\\
\hline
${}^{27}_{13}$Al & 0.0161 &  0.0173 & 0.0362 & 0.7054\\
\hline  
${}^{197}_{\phantom{1}79}$Au&  0.0974 & 0.146 & 0.189 & 13.07 \\ \hline
\end{tabular}
\end{center}
\caption{\label{tab:inputs-kitano} Reference values of nuclear overlap integrals
 and capture rates used in the phenomenological
analysis~\cite{Kitano:2002mt}. The overlap integrals are those
computed using ``Method I'' in Ref.~\cite{Kitano:2002mt}.  }
\end{table}

We use the results of Kitano et al.~\cite{Kitano:2002mt} 
on the rate of $\mu$-to-$e$ conversion in various nuclei for the
numerical results in this paper. In this appendix we summarize the
results from   Ref.~\cite{Kitano:2002mt} that we use in this paper. 
In terms of effective couplings $\tilde
g_{LV}^{(p,n)}$, $A_R$, and of nuclear overlap integrals $D$,
$V^{(p,n)}$, the branching ratio for the radiative decay is 
\beq
B_{\mu \to e \gamma} = 384 \pi^2   |A_R|^2  \ , 
\eeq 
and the branching ratio for conversion in nuclei is 
\beq 
B_{\mu \to e} = \frac{2 G_F^2m_\mu^5}{\Gamma_{\rm capt}}  
| A_R^*  \, D + \tilde{g}_{LV}^{(p)} \, V^{(p)} +
\tilde{g}_{LV}^{(n)} \, V^{(n)} |^2
\eeq 
The numerical values of the nuclear overlap integrals used in this paper
are given in Table~\ref{tab:inputs-kitano}.

The relations between the effective  couplings $\tilde
g_{LV}^{(p,n)}$, $A_R$, and the coefficients in the effective
Lagrangian, Eq.~(\ref{eq:eff-lag}), are:
\begin{align}
\tilde g_{LV}^{(p)} &= - \frac{4 v^2}{\LLFVsq} \, \Delta_{\mu e}^* \, 
\left[
(\frac14-s_w^2)(c_{LL}^{(1)}+c_{LL}^{(2)})+\frac32c_{LL}^{(3)}+c_{LL}^{(4u)}
+\frac12c_{LL}^{(4d)}-\frac12c_{LL}^{(5)}\right]\\
\tilde g_{LV}^{(n)} &= - \frac{4 v^2}{\LLFVsq} \, \Delta_{\mu e}^* \, 
\left[
-\frac14(c_{LL}^{(1)}+c_{LL}^{(2)})+\frac32c_{LL}^{(3)}+\frac12c_{LL}^{(4u)}
+c_{LL}^{(4d)}+\frac12c_{LL}^{(5)}\right]\\
A_R  &=\frac{e v^2}{\LLFVsq} \, \Delta_{\mu e} 
\left[- c_{RL}^{(1)}+  c_{RL}^{(2)}\right]
\end{align}

\end{document}